\documentclass{ws-procs9x6-cpt22}
\begin{document}

\newcommand{\refeq}[1]{(\ref{#1})}
\def\etal {{\it et al.}}

\title{SME Gravity in the Early Universe}

\author{N.A.Nilsson}

\address{Center for Quantum Spacetime,\\
Sogang University, Seoul, 121-742, Korea}

\begin{abstract}
In this talk, we give a short overview of recent work on cosmological solutions within the SME gravitational sector under the assumption of explicit spacetime-symmetry breaking. We show that for the special case of timelike diffeomorphism breaking, the resulting Friedmann equations can be written as standard FLRW cosmology with added dynamical dark energy, and we discuss primordial gravitational waves in the context of this model.
\end{abstract}

\bodymatter

\section{Introduction}
Motivated by the possibility of Planck-scale departures from known physics in the form of spacetime-symmetry breaking, a substantial research effort has been put towards highly accurate tests of Lorentz and CPT symmetry in gravity.\cite{alansamuel,datatables} By using the test framework called the Standard-Model Extension (SME), sensitive tests in the weak-gravity limit have been devised, for example within the Solar System, short-range gravity, gravitational waves, and more.\cite{solarsystem, shortrange, gws} Also, some specific models of spacetime-symmetry breaking have been studied in the context of gravity as well as future astrophysical observations.\cite{othertypes}

\section{Test framework}
The SME is an effective field theory, constructed to contain known physics as well as all possible combinations of Lorentz and CPT-violating terms to arbitrary order. The individual terms are constructed from indexed objects known as the {\it SME coefficients} contracted with conventional field operators; these coefficients control the magnitude of the symmetry breaking, and can be thought of as being implicitly suppressed by increasing orders of the Planck mass $M_{\rm Pl}$. These terms are invariant under observer Lorentz transformations but break particle invariance with respect to local Lorentz and diffeomorphism symmetry\footnote{In general, spacetime symmetries are broken when gravitational fields couple to fixed background tensors; a generic background field $X^{a,\hdots}$ defined in a local frame prescribes a preferred direction, and does \emph{not} transform covariantly under local \emph{particle} Lorentz transformations. On the level of the spacetime manifold, the background $X^{\alpha,\hdots}$ can instead break diffeomorphism invariance; an object defined in a local frame is connected to the spacetime manifold by means of a \emph{background} vierbein. 
}.\cite{alangravity} The minimal gravity sector truncates the series expansion at mass dimension $d=4$, serving as the leading order corrections to known physics, which in the gravity sector reads
\begin{equation}
    \mathcal{L} \sim R +(k_R)_{\alpha\beta\mu\nu}R^{\alpha\beta\mu\nu} + \mathcal{L}^\prime,
\end{equation}
where $(k_R)_{\alpha\beta\mu\nu} \rightarrow -uR +s^{\mu\nu}R^{(T)}_{\mu\nu}+t^{\alpha\beta\mu\nu}W_{\alpha\beta\mu\nu}$, $R^{(T)}_{\mu\nu}$ is the trace-free Ricci tensor, $W_{\alpha\beta\mu\nu}$ is the Weyl tensor, and $\mathcal{L}^\prime$ contains any dynamical terms of the SME coefficients. 
\section{Cosmological aspects of the SME}
In pure SME gravity, implications of {\it spontaneous} symmetry breaking at the time of inflation were studied, and it was found that the symmetry-breaking terms would generate anisotropies during inflation, which could subsequently leave imprints in the B-mode polarisation of the Cosmic Microwave Background (CMB); this lead to a constraint on the pertinent SME coefficients of $\leq 10^{-43}$.\cite{bonder} More recently, cosmological solutions were found in the context of explicit breaking, where the authors classified the properties of several accelerating solutions found in the 3+1 formulation of the SME.\cite{reyes}

\begin{figure}
\begin{center}
\includegraphics[width=3.0in]{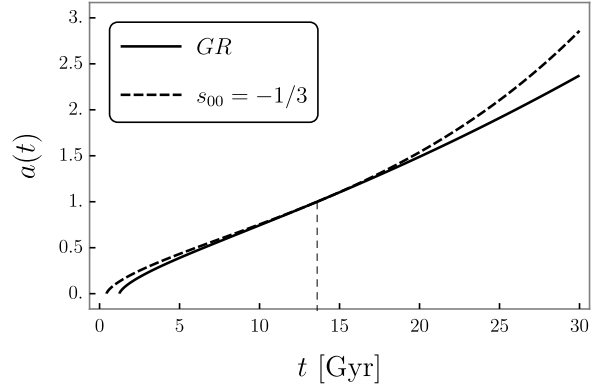}
\end{center}
\caption{Behaviour of the scale factor for a large negative value of the SME coefficient $s_{00}$ compared to the GR case.\cite{admsme}}
\label{fig:1}
\end{figure}

In Ref. \refcite{admsme}, the authors developed the 3+1 formulation of the SME in the case of explicit breaking and provided some of the first cosmological solutions; here, it was also found that a modified continuity equation for matter is necessary to avoid fourth-order time derivatives.\cite{admsme} Keeping only $s_{\mu\nu} = \text{diag}(s_{00},0,0,0)$ the Friedmann equation reads
\begin{equation}
    \frac{H^2}{H_0^2} = \Omega_m^0 a^{-3}+\Omega_r^0 a^{-4\eta_r}+\Omega_\Lambda^0 a^{-\eta_\Lambda}+\Omega_k^0 a^{-2}, 
\end{equation}
where $\eta_r = (4-3s_{00})/(4-2s_{00})$ and $\eta_\Lambda = 6s_{00}/(2-5s_{00})$. The radiation contribution can be expanded in $s_{00}$, giving
\begin{equation}
  \Omega_r^0 a^{-4\eta_r} = \Omega_r^0 a^{-4} + \Omega_r^0 a^{-4}\left[ \ln{(a)} s_{00} + \tfrac{1}{2}\ln{(a)}(1+\ln{(a)})s_{00}^2\right] + \mathcal{O}(s_{00}^3),
\end{equation}
which separates out the SME contribution to the evolution of $\Omega_r^0$. In this way, we can see that the SME pieces of the Friedmann equation behave as the LCDM model with added dynamical dark energy. Figure~\ref{fig:1} shows this behaviour for a large negative value of $s_{00}$.
\subsection{Primordial gravitational waves}
Taking the simple case of constant non-zero $s_{00}$ only, and introducing tensor perturbations around flat FLRW $ds^2 = a^2(d\eta^2+(\delta_{ij}+h_{ij})dx^idx^j)$, the quadratic SME gravity action reads
\begin{equation}
    \delta^2S=\tfrac{1}{4}\int d\eta d^3x((a^2-s_{00})h_{ij}^\prime h^{\prime~ij}-a^2\partial_kh_{ij}\partial^kh^{ij}),
\end{equation}
where $h_{ij}$ is a transverse-traceless tensor perturbation, $\eta$ is the conformal time coordinate, and $X^\prime \equiv \partial_\eta X$.\cite{pgwsme} Decomposing the tensor perturbations into the standard Fourier modes, the corresponding equations of motion read
\begin{equation}
    h_\lambda^{\prime\prime}(k)+2\Theta h_\lambda^\prime(k)+\Theta k^2h_\lambda(k) = 0,
\end{equation}
where $\Theta=a^2/(a^2-s_{00})$ will be different in each epoch. Here $\lambda=\{+,\times\}$ are the polarisations of the gravitational waves. An important consequence of the modified continuity equation in this model is that {\it there is no pure de Sitter or radiation-domination phase}, which affects the early-Universe dynamics. Instead, the conformal Hubble parameter ($\mathcal{H}=a^\prime/a$) in the resulting ``quasi de Sitter'' and radiation phases read
\begin{equation}
    \mathcal{H}_{\rm dS}=\frac{-2}{(2-\eta_\lambda)\,\eta},\quad \mathcal{H}_{\rm RD}=\frac{1}{(2\eta_r-1)\eta}
\end{equation}
where standard de Sitter and radiation domination are obtained for $s_{00}\rightarrow 0$. An interesting result appears when solving the equations of motion during radiation domination; since the comoving wave number $k$ can not be completely eliminated with the coordinate transformation $z=k\eta$, setting $k=1$ leads to a constraint on $s_{00}$ which matches the numerical value of CMB temperature, $s_{00}\leq T_{\rm CMB}/(1K)\cdot 10^{-9}$, for $T_{\rm CMB}=2.7255 K$, by virtue of the numerical solutions breaking down for larger values. 

The choice of constant $s_{00}$ is a toy model; more realistic choices of SME coefficients will surely describe early-Universe perturbations more accurately in the near future.

\section*{Acknowledgments}
NAN acknowledges support from
United Kingdom Research and Innovation (UKRI) and
the Basic Science Research Program through the National Research Foundation of Korea (NRF) funded by the Ministry of Education, Science and Technology
(2020R1A2C1010372).

\end{document}